\documentclass[12pt,preprint]{aastex}
\slugcomment{ver.Aug.08; re-submitted to AJ}
\slugcomment{ver.Sep.20; accepted to AJ}
\slugcomment{ver.Oct.12; accepted to AJ}

\shorttitle{sds3}
\shortauthors{Kashikawa et al.}

\begin{document}


\title{Subaru Deep Survey III. Evolution of Rest-Frame Luminosity Functions Based on the Photometric Redshifts for A $K'$-band Selected Galaxy Sample\altaffilmark{1} }


\author{Nobunari Kashikawa\altaffilmark{2,3}, Tadafumi Takata\altaffilmark{4}, Youichi Ohyama\altaffilmark{4}, Michitoshi Yoshida\altaffilmark{2,5}, Toshinori Maihara\altaffilmark{6}, Fumihide Iwamuro\altaffilmark{6}, Kentaro Motohara\altaffilmark{7}, Tomonori Totani\altaffilmark{8,9}, Masahiro Nagashima\altaffilmark{9}, Kazuhiro Shimasaku\altaffilmark{10,11}, Hisanori Furusawa\altaffilmark{4}, Masami Ouchi\altaffilmark{10}, Masafumi Yagi\altaffilmark{2}, Sadanori Okamura\altaffilmark{10,11}, Masanori Iye\altaffilmark{2,3}, Toshiyuki Sasaki\altaffilmark{4}, George Kosugi\altaffilmark{4}, Kentaro Aoki\altaffilmark{4}, Fumiaki Nakata\altaffilmark{10}}
\email{kashik@naoj.org}


\altaffiltext{1}{Based on data collected at Subaru Telescope, which is operated by the National Astronomical Observatory of Japan.}
\altaffiltext{2}{Optical and Infrared Astronomy Division, National Astronomical Observatory, Mitaka, Tokyo 181-8588, Japan}
\altaffiltext{3}{Department of Astronomy, School of Science, Graduate University for Advanced Studies, Mitaka, Tokyo, 181-8588, Japan}
\altaffiltext{4}{Subaru Telescope, National Astronomical Observatory, 650 N. Aohoku Place, Hilo, HI 96720, USA}
\altaffiltext{5}{Okayama Astrophysical Observatory, National Astronomical Observatory, Kamogata, Okayama 719-0232, Japan}
\altaffiltext{6}{Department of Astronomy, Kyoto University, Kitashirakawa, Kyoto 606-8502, Japan}
\altaffiltext{7}{Institute of Astronomy, University of Tokyo, Mitaka, Tokyo 181-8588, Japan}
\altaffiltext{8}{Princeton University Observatory, Peyton Hall, Princeton, NJ08544, USA}
\altaffiltext{9}{Theoretical Astrophysics Division, National Astronomical Observatory, Mitaka, Tokyo 181-8588, Japan}
\altaffiltext{10}{Department of Astronomy, University of Tokyo, Hongo, Tokyo 113-0033, Japan}
\altaffiltext{11}{Research Center for the Early Universe, University of Tokyo, Hongo, Tokyo 113-0033, Japan}


\begin{abstract}
We have constructed a very deep $K^\prime$-selected multicolor $BVRIz^\prime JK^\prime$ sample of $439$ field galaxies.
Based on this multicolor data, a photometric redshift for each sample galaxy was estimated.
The overall redshift distribution N(z) for the $K^\prime \leq 21.0$ sample is consistent with previous observations, and for the first time we derive N(z) down to $K^\prime=24.0$.
After taking account of the dust extinction and selection effects of the sample, the observed N(z) distribution is well described with the PLE model, while the hierarchical galaxy formation model shows an apparent deficiency of galaxies especially at $z\gtrsim2$.
The photometric redshift and the best-fit SED model evaluations allow the derivation of the rest-frame $K^\prime$, $B$, and UV($2000$\AA)-band luminosity functions and their evolutions.
The rest $K^\prime$-band luminosity function shows almost no evolution up to $z=3$, while the rest $B$ luminosity function shows mild luminosity evolution, and the rest UV luminosity function shows strong luminosity evolution.
These findings seem to be qualitatively in favor of the PLE model.
No evolution in the rest $K^\prime$-band luminosity function can also be consistent with the hierarchical galaxy formation model if $M/L_K$ decreases with redshift, that is, if the number density of $K^\prime$-band luminosity- selected galaxies does not significantly change with redshift while the number density of stellar mass-selected galaxies decreases with look-back time.
This trend corresponds to the evolution of the rest UV$(2000$\AA$)-K^\prime$ color, which gets bluer with increasing redshift.
We also found that more massive galaxies are redder in this rest-frame color in every epoch.
The rest-UV luminosity function of our $K^\prime$-selected galaxies shows a much shallower faint end slope at $z=3$ than that of previous estimations for rest-UV selected Lyman break galaxies.
As a consequence, the contribution to the global star formation rate of our $K^\prime$-selected galaxies is about $42\%$ of that derived from the integration of luminosity function of Lyman break galaxies at $z=3$.
This result suggests that a large fraction of the star formation rate density at $z>1.5$ comes from the contribution from the faint ($M_{2000}>-20$) blue galaxy population at high redshift universe that have not yet obviously been identified.

\end{abstract}




\section{Introduction}

The large galaxy sample with redshift determined to faint magnitudes allows us to understand the evolution of galaxies.
Previous deep spectroscopic surveys have shown several evolutionary trends for galaxies at $0\lesssim z\lesssim 1$, especially for the evolution of luminosity functions (LFs), which is one of the fundamental observational tools to trace the galaxy evolution.
\citet{lil95} have shown that the LF for red galaxies does not evolve significantly over $0\lesssim z\lesssim 1$, which infers that red massive galaxies must have already been assembled by $z=1$.
On the other hand, their blue subsample is evolving, which indicates that they are formed later than the red subsample.
\citet{cow96} have shown the evidence of the ^^ ^^ downsizing" trend in which more massive galaxies appear to be forming at higher redshifts.

Although it is important to see these evolutionary trends in the earlier epoch beyond $z=1$, it is difficult to construct a large galaxy sample in the familiar optical wavelength except for the distinct high z star-forming populations of Lyman break galaxies and/or Ly$\alpha$ emitters at $z\gtrsim 3$ \citep[Paper II]{ste99,ouc03}.
There are some possible interpretations of such luminous high z galaxies such as a core of today's massive $L>L^*$ galaxies \citep{ste96} or as a low-mass starbursting building blocks of the more massive galaxies seen today \citep{low97}.
However, it remains uncertain how these $z>3$ galaxies are related to present-day galaxies.

One of the explicit difficulties in tracing systematically the evolution of galaxies is that a single observed frame does not correspond to a single rest frame over wide redshift range, say $0\lesssim z\lesssim 3$.
The rest-optical frame in which we have seen the evolution trend at $0\lesssim z\lesssim 1$ shifts to the NIR region at $z\gtrsim 1.5$.
An observation in wide wavelength range is undoubtedly required to know the SEDs of galaxies and follow their evolution based on the same rest frame in which we can draw almost similar stellar contents at any redshifts.


There are two approaches to probe the rest-frame luminosity of galaxies. 
One is to change the observed band accordingly and follow the redshifting rest frame. 
This straightforward approach was taken by many studies.
\citet{cow99} traced the rest-$B$ band up to $z=1$ by $UBVI$ multiband observation.
\citet[Paper IV]{fur03} used multicolor $BVRi^\prime z^\prime$ bands and traced the rest $B$ band LF up to $z=1.25$.
NIR photometry for Lyman break galaxies enables them to derive their rest $V$-band luminosity at $z=3$ \citep{sha01}.
This approach gives a reliable way to trace the rest frame directly, though the study is restricted to a single rest band.

The other approach is to utilize the best-fit SED templates determined on photometric redshift method \citep{saw97,rud01}.
As have been already mentioned by many authors, the photometric redshift technique has less precision for redshift estimates than the spectroscopic redshift. 
However, spectroscopic redshifts are practical only for bright objects. 
Photometric redshifts can be extensively applied to fainter objects with a fair degree of confidence. 
They are also of great use in constructing a statisticaly large sample in the redshift range between $z=1.5$ and $z=2.5$, where redshifts can be identified only by NIR spectroscopy.
It works predominantly as a statistical technique rather than as a constraint on parameters that determine the SED of an individual object. 
Studies based on the photometric redshifts and those based on the spectroscopic redshifts complement each other.
If the SED of each sample galaxy in wide wavelength is given, the rest-frame luminosity of the galaxy can be estimated.
This method has the advantage to deriving the rest-frame luminosity at any redshift and at any wavelength, though the estimated luminosity inherits an uncertainty in determining the photometric redshift.


In this study, we have constructed a very deep $K^\prime$ band-selected sample of galaxies.
We then adopt the second approach described above with photometric redshifts derived from multicolor $BVRIz^\prime JK^\prime$ deep images and investigate the evolution of the LFs in the three ($K^\prime$, $B$, and UV) rest-frame bands out to $z=3.5$ where the $K^\prime$-band still traces the rest wavelength longer than an age-sensitive spectral break at $4000$\AA.
NIR band-selected samples may provide significant advantages over optical band-selected samples in studying galaxy evolution due to smaller extinction by dust and less type-dependent $k-$corrections \citep{man01} at these wavelengths.
Moreover, NIR selection provides samples that are not biased towards star-forming galaxies and allows estimation of the mass of galaxies over a wide range of redshift \citep{sar01}.
NIR photometry in general improves the estimate of photometric redshift (e.g., \citealp{bol00}).

The Subaru Deep Field (SDF) was imaged quite deeply at near-infrared with wide field near infrared camera CISCO and yielded some of the faintest galaxies ever observed, down to magnitudes of $J=25.5$ and $K^\prime=24.5$ \citep[Paper I]{mai01}, which is much deeper than most of previous $K^\prime$-band surveys (e.g., \citealp{cow96,bar99,fon00,rud01}).
The field was well selected, taking care to avoid large Galactic extinction and nearby clusters of galaxies and to get smaller airmass than the HDF at Mauna Kea (Paper I). 
The following optical imaging enabled us to estimate the photometric redshifts of these faint objects and offered the discovery of faint high z object candidates.
In the SDF, we have already evaluated the number counts in $J$ and $K^\prime$ bands, the near-infrared color distribution, and contribution to extragalactic background light (Paper I, see also \citealp{tot01a}).
\citet{tot01b} have investigated a detailed comparison of the number counts, colors, and size distribution for the SDF galaxies with theoretical galaxy formation models.
\citet{nag03} have also analyzed the SDF galaxies by using a semianalytic model of galaxy formation based on the CDM model.
In this paper, we will compare the redshift distribution with their models.

This paper will be organized as follows.
We present the process to construct the $K^\prime$-limited sample of SDF galaxies and estimating photometric redshifts as well as their spectroscopic calibration in \S 2.
In \S 3, we show the redshift distribution of our sample and compare it with galaxy formation models.
We discuss the evolution of LFs on rest $K^\prime$, $B$, and UV(2000\AA) bands in \S 4 and the evolution of rest UV$-K^\prime$ color in \S 5.
We will also describe in \S 6 the evaluation of star formation rate density (SFRD) derived from our LF estimate.
We give a summary in \S 7.
Throughout this paper we assume a flat, matter dominated universe ($\Omega_m=1$) with $H_0=75$kms$^{-1}$Mpc$^{-1}$ to compare with other major previous observations.

\section{$K^\prime$ band-selected Galaxy Sample} \label{datared}

\subsection{Data}

We have used a deep multicolor data set of $BVRIz^{\prime}JK^{\prime}$ images taken with the Subaru telescope on the Subaru Deep Field (SDF).
The SDF is centered on $\alpha=13^h 24^m 21.^s38; \delta=+27^{\circ} 29^{\prime} 23.^{\prime \prime}0 (J2000)$.
The $J$ and $K^\prime$ imagings and data reductions are described in the previous paper (Paper I).
The optical follow-up imaging observations were carried out during the commissioning runs of the FOCAS \citep{kas02} and Suprime-Cam \citep{miy98} in $2001$ February to June.
In this study, we used the Johnson $B$-, Kron-Cousins $R$- and $I$-band images taken with FOCAS and Johnson $V$- and SDSS $z^\prime$- band images taken with Suprime-Cam.
The observation and data reduction of the $V$ and $z^\prime$ data, which are one of the wide field imaging data for the SDF, will be described in coming paper \citep[Paper IV]{fur03}.
Here we briefly describe the imaging data of $B$, $R$, and $I$ taken with FOCAS.

FOCAS has a field of view as large as $6^\prime\phi$, which could entirely cover not only the SDF $2^\prime \times 2^\prime$ itself but also its flanking fields.
The SDF has flanking fields $25$ arcmin$^2$ in total size surrounding the SDF, though their limiting magnitude in $K^\prime$ ($K^\prime \leq 20.5$) is fairly shallower than that of the central field.
FOCAS has a couple of $2048\times4096$ pixel CCD chips with the pixel scale $0.103$ arcsec/pix.
The typical seeing size was $0.6-0.8$ arcsec. 
The total integration time was $2.5$hr, $4$hr, and $1$hr for $B$, $R$, and $I$ bands respectively.

The $B$, $R$, and $I$ images were reduced in a standard manner.
After flat-fielding, we have corrected the image distortion caused by the optics of FOCAS and the telescope.
The undistorted images have accuracy of $1.0$ pix rms.
This distortion pattern has been confirmed to be independent of the flexure of the telescope and the instrument.
The image shift caused by the flexure was about $0.6$ pix within the maximum exposure time of $30$min.
Relative sensitivity of two CCDs was corrected based on the difference of sky brightness assuming that the sky is uniform over the field of view.
Actually, the relative ratios of sky brightness of two CCDs on the same exposure were stable with $1\%$ fluctuation.
The sky was determined by fitting the unvignetted area of each frame with two-dimensional $3$rd or $4$th order polynomials and subtracted.
The relative shifts and rotations among dithered frames were measured and fixed using the positions of several point sources.
Then they were stacked each other using the IMCOMBINE task in IRAF.
Magnitude zero points were calibrated to the Johnson-Cousins system using Landolt photometric standard stars.
The accuracy of the zero points is estimated to be in the range $0.03-0.08$mag.
The magnitudes have also been corrected for the Galactic absorption, using $E(B-V)=0.018$ derived from \citet{sch98}, yielding $A_B=0.074$mag, $A_V=0.056$ mag, $A_R=0.046$ mag, $A_I=0.033$ mag, $A_{z^\prime}=0.025$mag, $A_J=0.015$ mag, and $A_{K^\prime}=0.006$ mag. 

The summary of optical imaging data used in this study is presented in Table~\ref{tbl-1}.
The $5\sigma$ limiting magnitudes in Vega system within $2.^{\prime \prime}0$ aperture for these images are $B\leq28.08$, $V\leq26.28$, $R\leq27.88$, $I\leq25.93$, $z'\leq24.27$, $J\leq25.11$, and $K^\prime\leq23.73$.

Among seven images, the $J$-band image has the smallest area, $2^\prime\times2^\prime$, on which the deep $K^\prime$-band imaging has carried out.
Therefore we first rescaled all the data to the same pixel scale ($0.^{\prime \prime} 103$ /pix) as the $B$, $R$, and $I$ data, corrected relative rotation based on the positions of several bright point sources, and extracted the $1.^\prime 90 \times 1.^\prime 97$ region common to the $J$- and $K^\prime$-band images for further analysis.
The error of this image translation was below $0.02$ arcsec.

\subsection{Object Detection and Photometry}

We extracted objects using SExtractor \citep{ber96}.
The objects were extracted from the $K^{\prime}$-band image as a reference image and photometric measurements were performed on the other image for the exact positions where sources were detected on the reference image. 
We choose a $4.0$pix FWHM Gaussian convolution kernel in object detection on the reference $K^\prime$-band image.
We used a detection threshold of $24.1$ mag/arcsec$^2$ corresponding to the $1.5\sigma$ above the background noise level in $K^\prime$ band, the same as in Paper I.

To measure colors of detected objects, aperture photometry was carried out for seeing-corrected images, which were created by convolving Gaussian filters to match the seeing size to that of the $V$ band, the largest among the seven images.
At first we used a fixed aperture. 
However, it was difficult to choose an adequate aperture size because photometry with a large aperture is affected by close neighboring objects while a small aperture does not cover the same physical sizes for both nearby objects and distant objects at the same time. 
Finally we chose an aperture size adjusted for each object size but kept identical for all seven images.    
To choose the adequate aperture size in deriving colors of objects, we have tried many times with different aperture sizes and confirmed the consistency between the resulting photometric redshifts and spectroscopic redshifts for calibration samples (see Section $2.3$).
We found that $2\times$FWHM (FWHM was measured on $K^\prime$-band image) aperture for each object is the optimum aperture size to derive correct photometric redshifts.
In this way, we have measured total magnitudes (^^ ^^ {\tt MAG$\_$BEST}" magnitutudes from SExtractor), $2\times$FWHM aperture magnitudes for all the seven images.
It should be noted that the $2\times$FWHM aperture magnitude is only used for SED fitting in deriving the photometric redshift, and further analysis is based on the total magnitude. 
The nine objects that have stellarity$>0.9$ with $K^\prime \leq 23.0$ were eliminated as Galactic stars from our galaxy sample.
We have also removed $12$ objects whose photometry is strongly affected by nearby bright stars in one of the seven band images.

Our final $K^\prime$-selected catalog contains $439$ objects with the magnitude limit $K^\prime_{tot}\leq 24.0$.
According to Paper I, the completeness is over $70\%$ at this limiting magnitude.

\subsection{Photometric Redshifts}

We used the publicly available $Hyperz$ code by \citet{bol00} to determine photometric redshifts for our seven-band $BVRIz'JK^\prime$ photometry.
The $Hyperz$ code derives a photometric redshift for each input multi-photometry set according to the standard SED fitting minimization procedure.
The detail of the method as well as discussions about the accuracy of evaluated redshifts are presented in \citet{bol00}, therefore we describe here briefly about our choices of major input parameters.
We used five (Burst, E, Sa, Sc, and Im) synthetic galaxy SED templates of Bruzal \& Charlot model (GISSEL98, \citealp{bc93}) to compare with the observed SED.
The default SED templates applied in $Hyperz$ assume the Miller \& Scalo IMF from $0.1M_\odot$ to $125M_\odot$, solar metallicity $Z=Z_\odot=0.02$, ages ranging from $10^6$yr to $2\times10^{10}$yr, and Madau (1995)'s formula for continuum depression due to the intergalactic absorption blueward of Lyman $\alpha$ emission.
The adopted SED templates are assumed for an instantaneous single-burst SFR model for ^^ ^^ Burst" SED, and exponentially decaying SFRs with $e$-folding timescales of $1$, $3$, $15$ Gyr and infinity ($i.e.$, constant SFR) to match the colors of present-day galaxies of different spectral types from ^^ ^^ E" to ^^ ^^ Im".
Though three more additional spectral type templates can be applied in $Hyperz$, the accuracy for photometric redshift $z_{phot}$ estimate was found to be unchanged when applying all the eight templates.
We have chosen a \citet{cal00}'s extinction law with $A_V$ ranging between $0.0$ and $1.8$.
The step in redshift in searching the solutions between $z=0$ and $z=7$ is $\Delta z=0.1$, though the resolution in the final minimization is $10$ times better.
The results are not significantly affected by the choice of smaller $\Delta z$.
We have also constrained on the age of the template to be less than the age of the universe at the redshift under consideration.
The transmission curve for a given filter set was defined based on the real measured curve for each instrument.  
An uncertainty of derived photometric redshift for each object was estimated from the $\chi^2$ probability distribution at the confidence level $90\%$ derived in SED template matching, which will be shown later.
As expected, the error becomes larger as $K^\prime$ magnitude goes fainter.


We examined the accuracy of our determination of photometric redshifts, $z_{phot}$, against spectroscopic redshifts, $z_{spec}$.
We have measured spectroscopic redshifts for $76$ galaxies in $5^\prime \times 5^\prime$ field centered on SDF.
All these spectroscopic data were taken during the commissioning runs of FOCAS MOS mode in April and May, $2001$.
The spectra cover $4700-9000$\AA~with a pixel resolution of $2.8$\AA~under the $0.^{\prime \prime} 8-1.^{\prime \prime} 0$ seeing size and were reduced in a standard manner.

\begin{figure}
\plotone{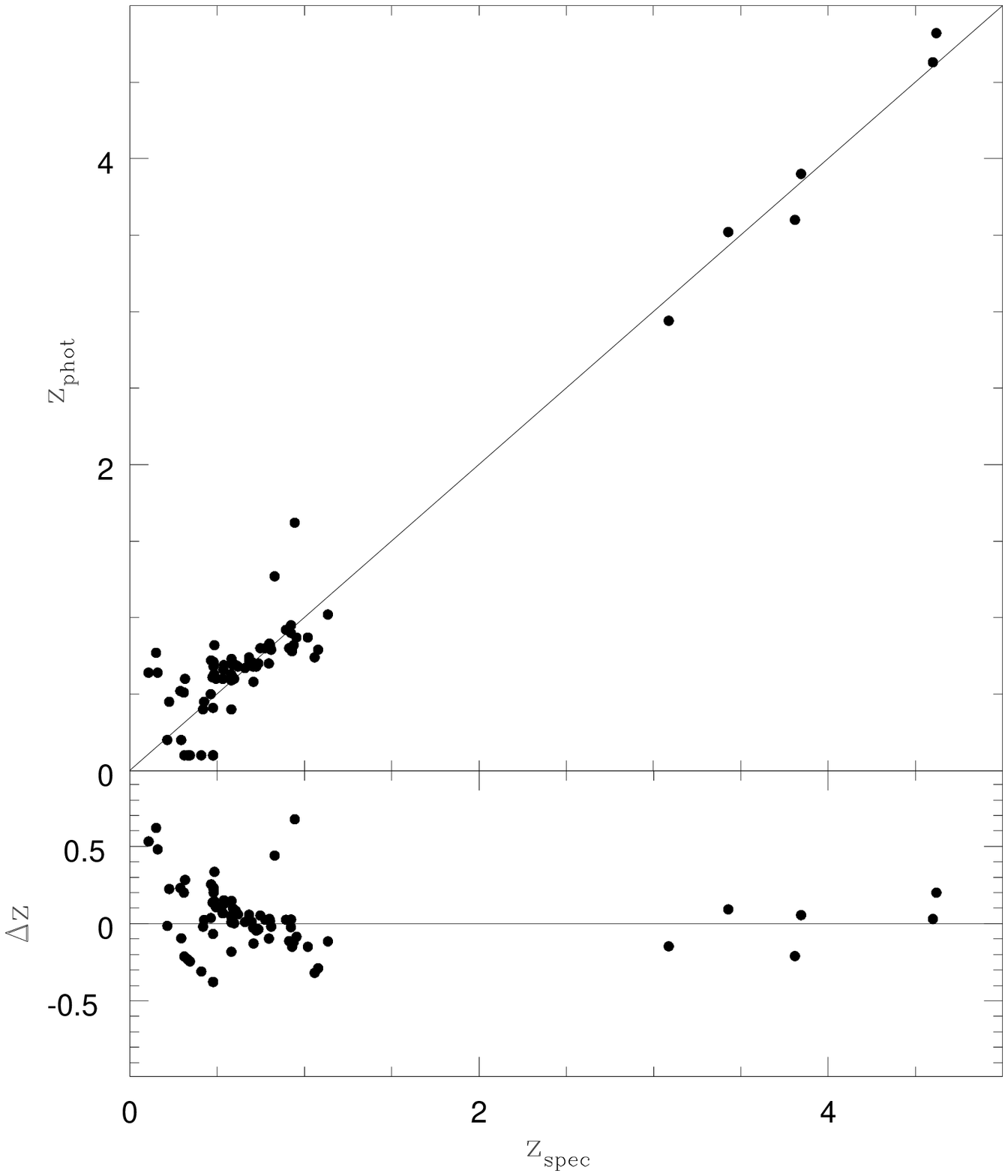}
\caption{Comparison between $z_{spec}$ and $z_{phot}$ for our spectroscopic sample of SDF.
\label{fig_zsp}}
\end{figure}

In this $5^\prime \times 5^\prime$ SDF flanking fields, $J$-band data are not available except for the central $2^\prime \times 2^\prime$ region and the $K^\prime$-band image has shallower limiting magnitudes, $K^\prime<20.5$.
Five galaxies of $76$ galaxies belong to the SDF central field in which $J$-band photometry is available.
We have confirmed an excellent agreement of the estimated $z_{phot}$ between the photometric data sets with and without $J$-band for these five galaxies. 
The comparison of $z_{phot}$ with $z_{spec}$ for our spectroscopic sample is shown in Figure~\ref{fig_zsp}, in which we found good agreement, though we don't have a large enough spectroscopic galaxy sample at $1.5<z$.
The mean error is $\Delta z= \langle z_{spec}-z_{phot} \rangle =-0.038$ with $\sigma=0.207$.
The large inaccuracy seen in $z \leq 0.4$ would be caused by the lack of $U-$band photometry which can detect a Balmer break at this redshift range.
If objects with $z<0.4$ are removed, the accuracy improves to $\sigma=0.178$.
Note that we don't have any apparent catastrophic errors, though the spectroscopic sample itself is small.
If five objects with $\Delta z >0.4$ are removed, the accuracy improves to $\sigma=0.160$.

\section{Redshift Distribution}

\begin{figure}
\plotone{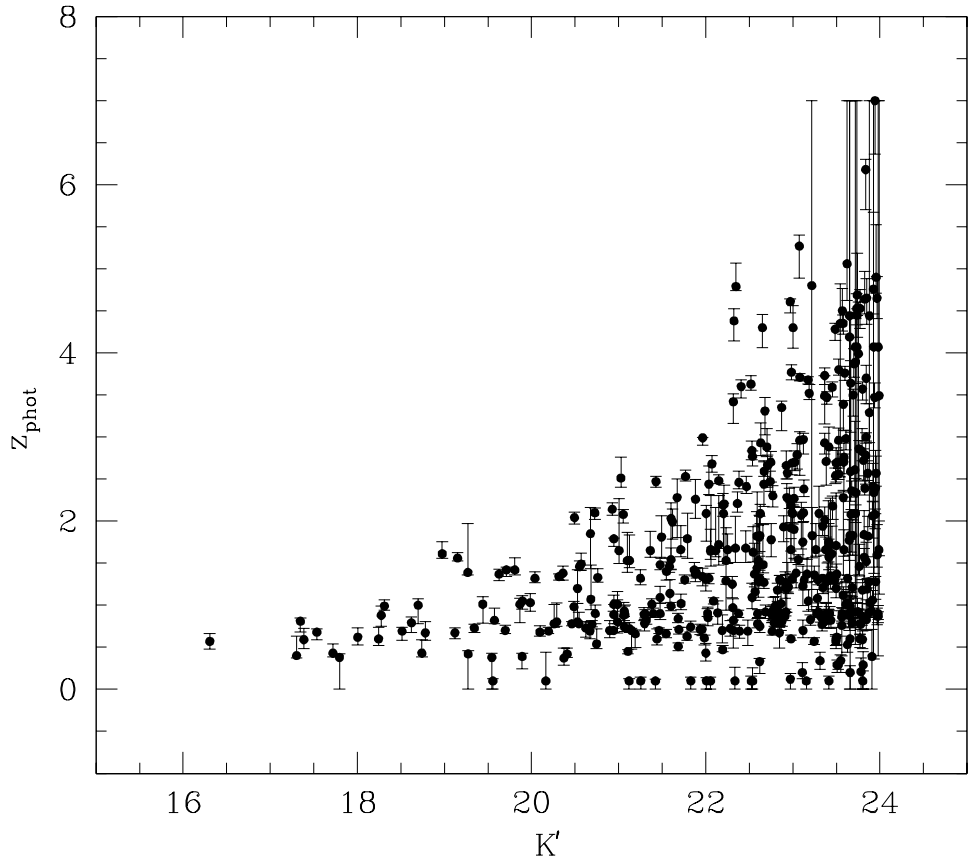}
\caption{$K^\prime$ magnitude vs. $z_{phot}$ diagram. The error bars of photometric redshifts are evaluated from the $\chi^2$ probability distribution at confidence level $90\%$. 
\label{fig_kz}}
\end{figure}

\begin{figure}
\plotone{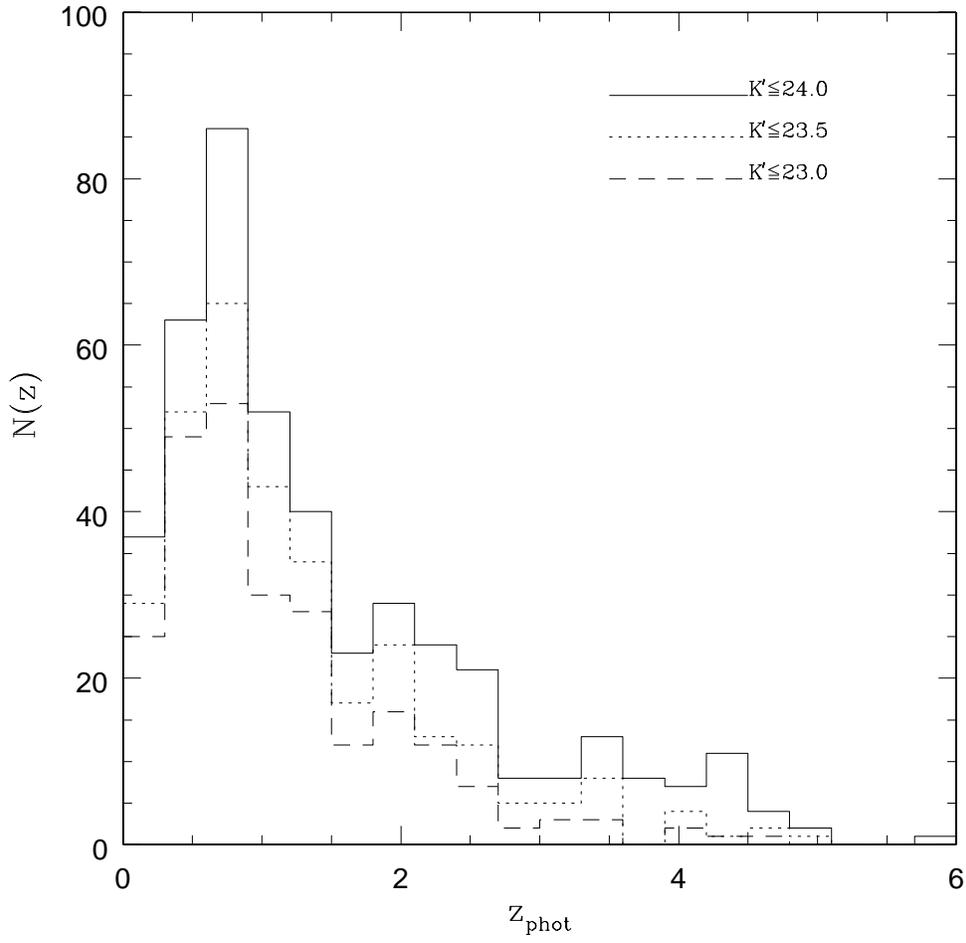}
\caption{Redshift distribution of our SDF sample with different limiting magnitudes of $K^\prime \leq 23.0$, $K^\prime \leq 23.5$, and $K^\prime \leq 24.0$. \label{fig_nz}}
\end{figure}

We describe in this section the derived redshift distributions of our sample.
Figure~\ref{fig_kz} shows the $K^\prime$-z relation.
The error bars of the determined $z_{phot}$ values are estimated from the $\chi^2$ probability distribution at confidence level $90\%$ derived in SED template matching.
Figure~\ref{fig_nz} shows the redshift distribution of our $K^\prime$-selected sample. 
The distribution has a peak around $z=0.8$ which is slightly larger than the N(z) peak around $z=0.5-0.6$ derived by \citet{fon99} or \citet{rud01} for $K\lesssim 21$ samples which include HDF-N,S fields.
However, if we restrict our sample to brighter galaxies having $K^\prime \leq 21.0$, the peak moves slightly toward a lower z at $z=0.6$.
The number density of galaxies at $1<z<1.5$ is $4.8$ galaxies/arcmin$^2$ at $K^\prime \leq 21.0$, which is consistent with the result of \citet{fon99} of $4.4$ galaxies/arcmin$^2$ at $K^\prime \leq 21.0$.
Thus, our N(z) is consistent with those of \citet{fon99} and \citet{rud01}.

We can use this overall redshift distribution of $K^\prime$ band-selected galaxies to compare the predictions of theoretical models of galaxy formation.
The redshift distribution for the $K$-selected galaxy sample was first pointed out by \citet[KC98]{kc98} as an effective tool to discriminate amoung galaxy formation models.
The hierarchical clustering model, in which massive galaxies were assembled much more recently through merging of small galaxies, predicts less galaxy abundance at high z ($z>1$) than that expected from the pure luminosity evolution (PLE) model.
The recent results of \citet{fon99} and \citet{rud01} for the $K^\prime \leq 21$ sample were both consistent with the hierarchical clustering model.
Since our $K^\prime$-selected sample is much deeper than \citet{fon99} and \citet{rud01} by $3$mag, the derived redshift distribution reflects both a high z ($z>2$) population and intrinsically faint population in the local universe and is crucial and useful in reconciling observation with theoretical models.
Actually, the fraction of galaxies at $z\geq 1.6$ detected in $K^\prime \leq 21$ sample of \citet{fon99} was only $9\%$.

\begin{figure}
\plotone{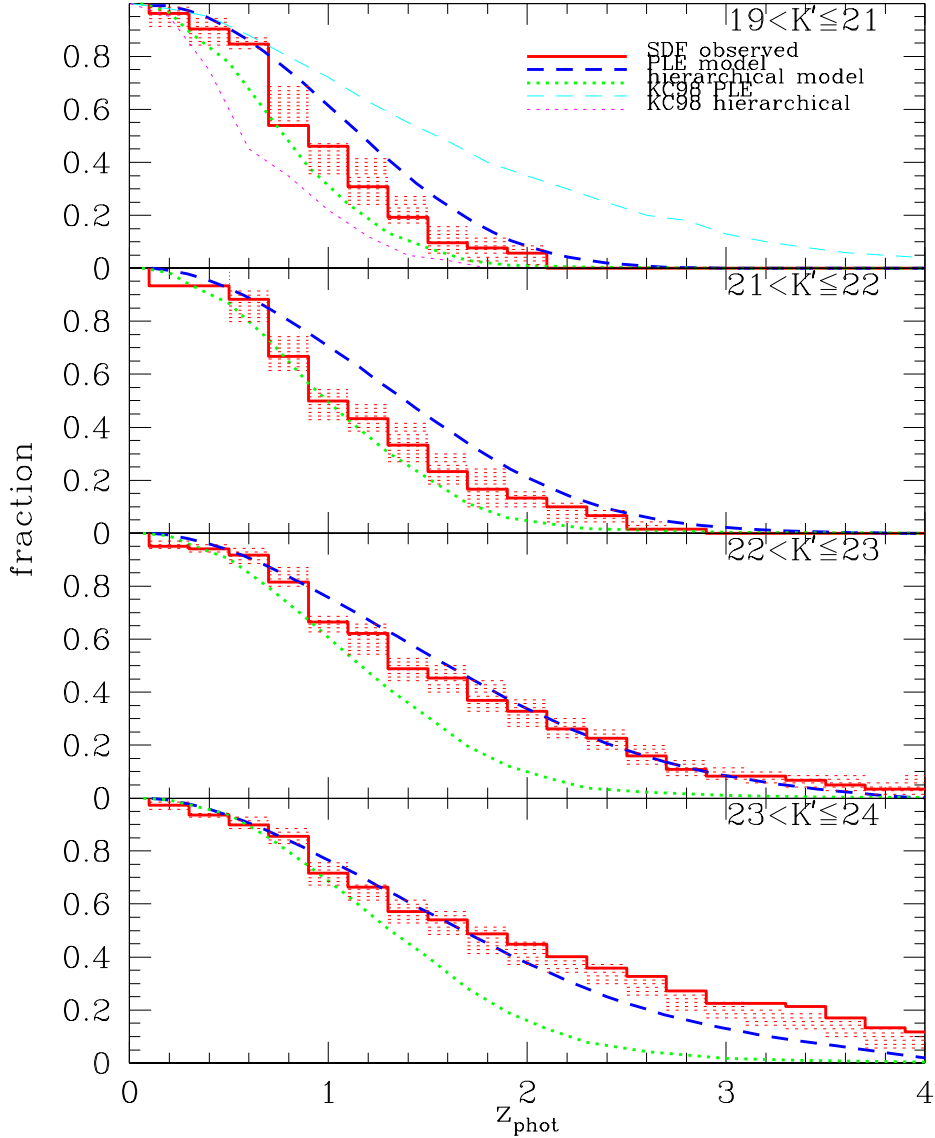}
\caption{{\it Top to bottom}: Normalized cumulative redshift distribution for our SDF sample for the bright-magnitude bin to the faint magnitude-bin.
The histograms show the results of observation.
The shaded regions show the $\pm3\sigma$ deviated counts estimated by Monte Carlo realizations when photometric redshift errors are taken into account.
The heavy dashed lines denote the prediction of the PLE model by Totani et al.(2001b), while the heavy dotted lines denote the the hierarchical galaxy formation model by Nagashima et al.(2003). 
{\it Top}: Prediction of PLE and hierarchical clustering models in KC98 as well.\label{fig_nzc}}
\end{figure}

Following the same manner as KC98 and previous observations, Figure~\ref{fig_nzc} demonstrates the normalized cumulative redshift histogram for our $K^\prime$-selected sample, as well as the theoretical predictions of both the PLE and hierarchical clustering models.
The thick solid histograms show the results of observation, while other lines show the model predictions.
We have carried out Monte Carlo realizations to see the distortion that the error of our derived photometric redshift could cause on the resulted N(z). 
At first, we have created a ^^ ^^ pseudo-galaxy-catalog" in which redshift of each galaxy was assigned to the photometric redshift solution plus a random error perturbed in the $90\%$ range of confidence interval in the SED $\chi^2$ fitting procedure (same as the error bars denoted in Figure~\ref{fig_kz}). 
In this process, magnitude of each galaxy was also assigned with a random error perturbed within the measured photometric error by SExtractor.
We assumed here that the error in photometric redshifts is independent of the photometric error, though a large photometric error would cause a large error of photometric redshift in most cases.
The redshift distribution is recalculated for this sample. 
We repeated the process of generating such pseudo-galaxy-catalogs $200$ times and investigated the scatter of derived redshift distribution.
The shaded regions in Figure~\ref{fig_nzc} show the $\pm3\sigma$ deviated counts estimated by this method.

The thick dashed lines show the PLE model by \citet{tot01b}.
The model normalizes its number density of galaxies at $z=0$ by the $B$ band luminosity function, and it probes the evolution backward in time.
Galaxies are classified into six morphological types of E/S0, Sab, Sbc, Scd, Sdm, and the distinct population of dwarf elliptical galaxies.
Their luminosity and SED evolutions are described by a standard galaxy evolution model of \citet{ari87} and \citet{ari92}.
All galaxies are simply assumed to be formed at a single redshift, $z_F=5$. 
A reasonable evaluation of dust extinction by a screen model, and selection effects such as cosmological dimming of surface brightness are taken into account.
The details of the model are described in \citet{yos93} and \citet{ty00}.
Their selection criteria for object detection are completely the same as ours and those in Paper I.
This PLE model, which takes account the two distict populations of giant elliptical galaxies and dwarf elliptical galaxies, was found to match extremely well with the observed $K$-band galaxy counts on the SDF beyond $K>22.5$ where the PLE model with a single elliptical population could not explain the observed galaxy abundance.
In their original model of \citet{tot01b} they have predicted the N(z) for the SDF based on an isophotal magnitude in which selection effects are easy to take into account.
We recalculated the N(z) predicted by the model based on the total magnitude in this study.

On the other hand, the thick dotted lines represent the semianalytical hierarchical galaxy formation model by \citet{nag03}.
This model is based on the structure formation theory in the CDM model, which corresponds to the hierarchical clustering of dark halos, and contains some important physical processes concerned with the evolution of the baryonic components such as gas cooling, star formation, supernova feedback and galaxy mergers.  
In this model, merging histories of dark halos are realized by using a Monte Carlo method assuming a power spectrum of initial density fluctuation in a Gaussian random field, and then, the above baryonic processes are followed in each dark halo.
Thus star formation histories, luminosities and colors of individual galaxies are predicted with this model.
The details of the model itself are presented in \citet{nag01} in which the model well reproduces the local luminosity function, the \ion{H}{1} gas mass fraction, galaxy sizes and so on.
The \citet{nag03} model employed in this study improves on the \citet{nag01} model by taking account almost the same selection effects as those in \citet{tot01b} and dust extinction, and can predict the number counts of galaxies consistent with that of observed on the SDF.
Note that the model of \citet{tot01b} was calculated for a cosmology with $\Omega_m=0.2$ and $\Omega_\lambda=0.8$ and that of \citet{nag03} was calculated with $\Omega_m=0.3$ and $\Omega_\lambda=0.7$ as were those of their original models, whereas an overall distribution of observed photometric redshift does not depend on cosmology.
In the brightest $19<K^\prime \leq 21$ bin, we also overlay the model predictions of the PLE and hierarchical clustering models of KC98 for comparison.

It should be noted that the hierarchical clustering model apparently shows a strong deficiency compared with the observed N(z), especially at high redshifts ($z>2$).
In the $22<K^\prime \leq 23$ bin, which contains $119$ sample galaxies and still attains the $100\%$ completeness, the PLE model agrees with the observed count better than hierarchical galaxy formation model does.
The PLE model can account for the abundance of the high-z ($z>2$) galaxies.
On the other hand, its underprediction of the high-z galaxy abundance as compared with that of the PLE model of KC98 is attributed to the dust evaluation in the model.

We performed a two-sample form of the Kolmogorov-Smirnov test (K-S test) to see the statistical agreement between observation and models.
The chance probability, as well as the $\chi^2$ values (in parentheses) are listed in Table~\ref{tbl-ks}.
The K-S test accepts at the $1\%$ significance level the hypothesis that the observation and the PLE model in the $19<K^\prime \leq 23$ bins are drawn from the same distribution, while the hierarchical galaxy formation model agrees with observation only in the bright $19<K^\prime \leq 22$ bins.
In the faintest $23<K^\prime \leq 24$ bin both the model predictions are found to be different from the observation.
In this bin, the observed count shows a high z tail which may be an effect of having more galaxies at high z and fewer galaxies at low z than in the models.
Though the error-shaded regions seem to be distributed below the observed counts, it is partly because the pseudo-galaxy catalog lacks the objects whose reassigned magnitudes are beyond the limiting magnitude $K^\prime=24.0$.
Note that the completeness of the sample is $100\%$ at $K^\prime=23.5$, though $70\%$ at $K^\prime=24.0$ (Paper I).
In the galaxy count, the model prediction begins to show a slight deviation from the observation at this faint magnitude \citep{tot01b,nag03}.
The discrepancy in this faintest bin is presumably caused by the large photometric errors which give incorrect solutions of photometric redshifts. 
More detailed refinements of the model may also be required.

As noted above, previous works on the KC98 test such as \citet{fon99} and \citet{rud01} found that their N(z) is well described with the hierarchical clustering model prediction and that the PLE model overpredicts the abundance at all redshifts.
Our observed N(z) itself is consistent with those observed by \citet{fon99} and \citet{rud01} at least at $K^\prime \leq 21$, and the KC98 test with our N(z) would give the same conclusion if we adopt the model predictions of KC98.
In this study, we introduce more realistic models of PLE and hierarchical galaxy formation that take account of both the selection effects of the sample and dust extinction, and found that the PLE model is favored by the observed N(z), while the hierarchical galaxy formation model prediction is substantially below the observed N(z) at high z. 
\citet{cim02} concluded that their spectroscopic redshift distribution at $K_s<20$ of the K20 survey is consistent with the PLE models that take into account the selection effects and dust extinction.
It might suggest that we need some additional physical processes in the CDM model.
We note, however, that the result is obtained on the very small field of the sky, therefore it should be confirmed by a future deep and wide-field $K$-selected sample.

\section{Evolution of Rest-frame Luminosity Functions}

Using the estimated photometric redshift catalog, we now investigate the galaxy population and its evolution by means of luminosity functions.
Our wide wavelength coverage and SED estimate for each sample galaxy allow us to derive LFs in the three rest-frame $K^\prime$, $B$, and UV($2000$\AA) bands.
We will investigate the evolution of rest-frame LFs at $z<3.5$, where $K^\prime$-band traces the rest-frame wavelength longer than $4000$\AA.
Though it is quite interesting to see the difference of rest-frame LFs for a blue or red subsample (e.g., \citealp{lil95}) or a spectral type (e.g., \citealp{fol99}) subsample, our sample is not so large as to be subdivided into small subsamples with enough statistical confidence.
Such a detailed analysis would be possible in the future when a large NIR imaging survey is carried out.

\subsection{Rest $K^\prime$ Frame Luminosity Function}

First, we derive the rest $K^\prime$-band luminosity functions.
For each galaxy we have determined a rest-frame absolute $K^\prime$ magnitude, $M_{K^\prime}$ as

\begin{equation}
M_{K^\prime}=m_{K^\prime}-5\mbox{log}(d_L/10\mbox{pc})-k(z),
\end{equation}

where $d_L$ is the luminosity distance and $m_{K^\prime}$ is the apparent $K^\prime$ magnitude.
The $k-$correction $k(z)$ can be derived from the best-fit SED template for each galaxy.

To determine the LFs we use a traditional $1/V_{max}$ method \citep{fel77},

\begin{equation}
\phi(M)dM=\frac{4\pi}{d\omega}\sum \left[ \int_{z_{min}}^{z_{max}}\frac{dV}{dz} dz \right] ^{-1},
\end{equation}

where $dV/dz$ is the comoving differential volume.
The limits $z_{min}$ and $z_{max}$ are defined as the smallest and the largest redshifts, respectively, that a given galaxy could have and still be included in a sample (see detail in \citealp{eli96}).
The sum is carried out over all galaxies in the sample lying within the specified range of redshift and magnitude.
The volume is based on an effective solid angle, d$\omega$, of the survey of $3.739$ arcmin$^2$.

\begin{figure}
\plotone{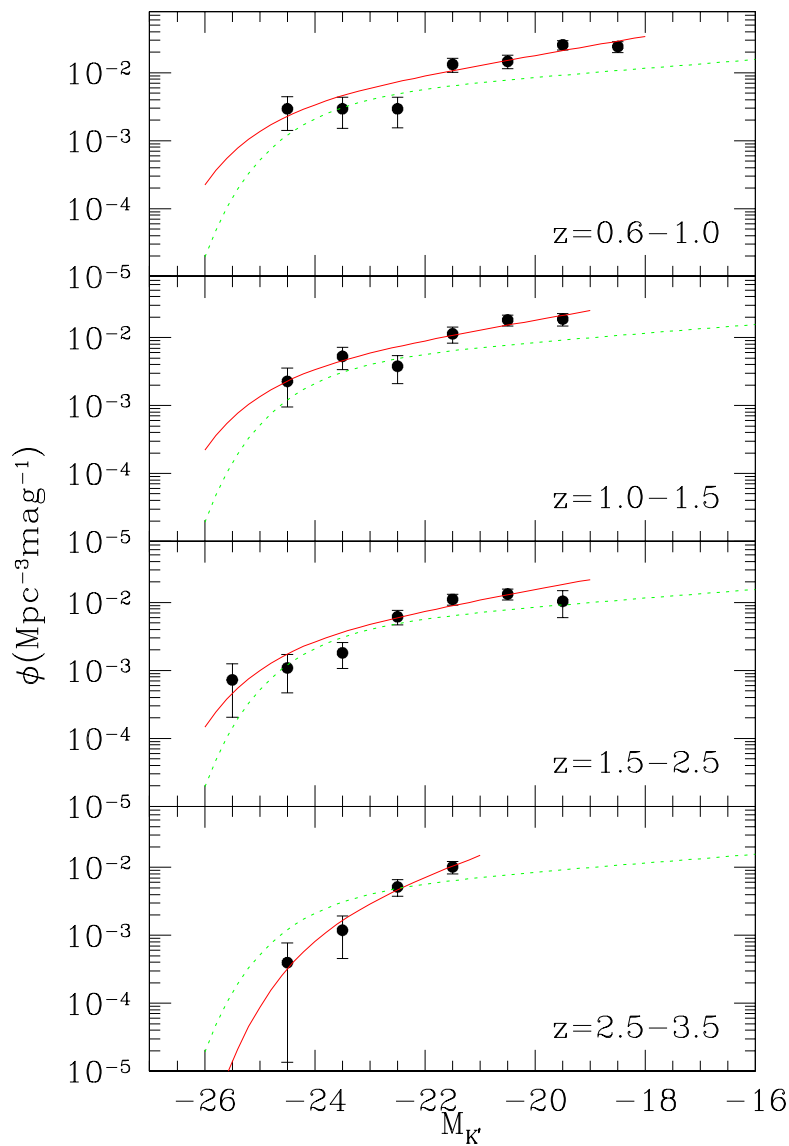}
\caption{{\it Top to bottom}: Rest $K^\prime$-band luminosity functions at $z=0.6-1.0$, $z=1.0-1.5$, $z=1.5-2.5$, and $z=2.5-3.5$.
Filled circles show the luminosity function determined in the present study, and solid lines are the best-fit Schechter functions. 
The dotted line shows the local $K^\prime$ luminosity function by Loveday (2000) \label{fig_lfk}}
\end{figure}

The rest $K^\prime$ frame LFs for four redshift intervals at $0.6<z \leq 3.5$ are shown in Figure~\ref{fig_lfk}.
We have neither enough sample galaxies to derive a reliable LF nor good accuracy in redshift identification at $z<0.6$, so we omit this redshift interval.
Instead, we refer the local $K$-band LF derived by \citet{lov00} for the Stromlo-APM Redshift Survey, shown as a dotted line on each frame in Figure~\ref{fig_lfk}.
Although Loveday's sample is smaller than that used in another local $K$-band LF of 2dF survey \citep{col01}, it extends to a magnitude as faint as $M_K=-16$, with smaller statistical error, which is adequate to compare with our sample.
The error bars show the statistical uncertainties evaluated from a bootstrap resampling method with a series of $500$ Monte Carlo simulations.
We fit the LF by a usual Schechter parameterization and list the best-fit parameters in Table~\ref{tbl_lf}.

The derived $K^\prime$ LFs at $0.6<z \leq 1.0$ differ slightly from the local one, having hints of a steeper faint-end slope and brighter bright end, though the overall shape of the LF seem to agree with that of local $K^\prime$-band LF taking account the normalization difference.
The most striking result perceived from this figure is that the $K^\prime$-band LF remains almost unchanged from $z=0.6$($0.0$) to $z=2.5$ and shows decline at the bright end at $2.5<z \leq 3.5$.
We have evaluated the robustness of our resulted LFs against the errors of our photometry and the photometric redshift estimate by the Monte Carlo realizations, which is as the same as those we performed in \S $3$.
The evaluated $1\sigma$ scatter of derived LF parameters, {\it i.e.}, $\alpha$, $M^*$, and $\phi^*$ are denoted in Table~\ref{tbl_lf}.
We found that the effects of photometric and redshift errors are small with negligible effect on $\alpha$, and at most $0.5$mag in $M^*$.
The $K^\prime$-band LF at $2.5<z \leq 3.5$ also has signs of a steeper faint end slope than that at lower z ones, which is expected in the context of the hierarchical clustering model. 
However, the trend may not be statistically significant because of the lack of enough sample objects in faint ($M_{K^\prime}>-21$) bins.

\citet{cow96} have derived the rest $K^\prime$-band LF for $0.6<z<1.0$ galaxies. 
It remains almost unchanged from that for $0<z<0.2$.
\citet{dro01} also showed no significant evolution of rest $K^\prime$-band LF for $z<1.2$.
\citet{lil95} have also noted that rest-frame red luminosity galaxies have a weak evolution up to $z=1$.
This trend has been found to continue up to $z\lesssim 3$ in this study.
The almost no evolution in the rest $K^\prime$-band LF implies that the old stellar population in galaxies was already in place by $z=2-3$ and that their stellar mass did not change significantly at $0<z<3$.
It seems that the present result is inconsistent with the hierarchical clustering model.
\citet{pap02} estimated the stellar masses for the HDF-N galaxies by using population-synthesis models and found that massive present-day galaxies were not fully assembled by $z\sim2.5$ as predicted by the hierarchical galaxy  formation model, which might be also inconsistent with ours.
However, \citet{dro01} have shown that the $M/L_{K}$ slightly decreases with redshift at $z<1.2$.
If this trend continues up to $z\sim2.5$, all the findings are consistent with the fact that the number density of $K^\prime$ band luminosity-selected galaxies does not remarkably change with redshift, while the number density of stellar mass-selected galaxies decreases with look-back time.

It should be noted that we are assuming here a rather large extrapolation along the SED template from the observed $K^\prime$-band to the rest $K^\prime$-band especially at $2.5<z \leq 3.5$. 
The reliability of this extrapolation depends on some appropriate assumptions in constructing the SED templates, based on our fundamental knowledge of star formation histories in a galaxy.
Recently, \citet{sha01} and \citet{pap02} have attempted to derive the stellar masses for individual sample of Lyman break galaxies at $z\sim3$. 
In these studies, their UV to near-IR photometries were fitted with a series of stellar population synthesis models, which is basically the same SED-fitting method as photometric redshift except that they considered carefully almost full allowable range of various input parameters in constructing SED templates. 
This constraint method can be applied only to the samples whose redshifts are reliably determined by spectroscopy. 
Although some interesting trends have been found among parameters of star formation histories, they found only loose constraints on each of the parameters including stellar masses. 
This means that the stellar mass measurement even at observed $K^\prime$ band at $z\sim3$ has some uncertainty.
The most significant uncertainty factor in deriving stellar mass is a second stellar population component \citep{pap02} of old stars that have formed in a burst that occurred before ongoing active star formation. 
The stellar masses with this two-component model are estimated to be $3-8$ times of that of the single-component models. 
While this old population, if any, actually influences on the estimate of stellar masses, its allowable flux contribution is supposed to be small because of its high mass-to-light ratio. 
For example in Figure $19$ of \citet{pap02}, which shows the effect of a second old stellar population on the galaxy SED, the second component increases the inferred rest $K^\prime$-band luminosity by a factor of less than $\sim1.4$, which corresponds to $\sim0.37$ mag brightening. 
As a consequence, our extrapolation of SED and derived rest $K^\prime$-band magnitudes at high z in this study could have uncertainty to the same extent.

In the following subsections, we will see the evolution of rest-frame LFs of bluer rest bands that trace young stellar population and are sensitive to star formation activity.

\subsection{Rest $B$ Frame Luminosity Function}

We have also estimated the rest $B$-band luminosity functions with almost the same kind of $K^\prime$ LF.
Six objects out of $439$ in the $K^\prime$-selected sample were not detected in the $B$-band image.
We have derived their inferred $B$-band magnitudes from their best-fit SED templates, convolved with a $B$-band response curve.
The consistency between photometric $B$-band magnitudes and these inferred $B$ magnitudes from their best-fit SEDs were checked with sample galaxies except these six $B$-dropout galaxies, and their agreement was found to be within $0.56$ mag rms.
The inferred $B$ magnitudes of the six $B$ dropout galaxies were all beyond the $1\sigma$ limiting magnitude $29.76$ of the $B$-band image.

\begin{figure}
\plotone{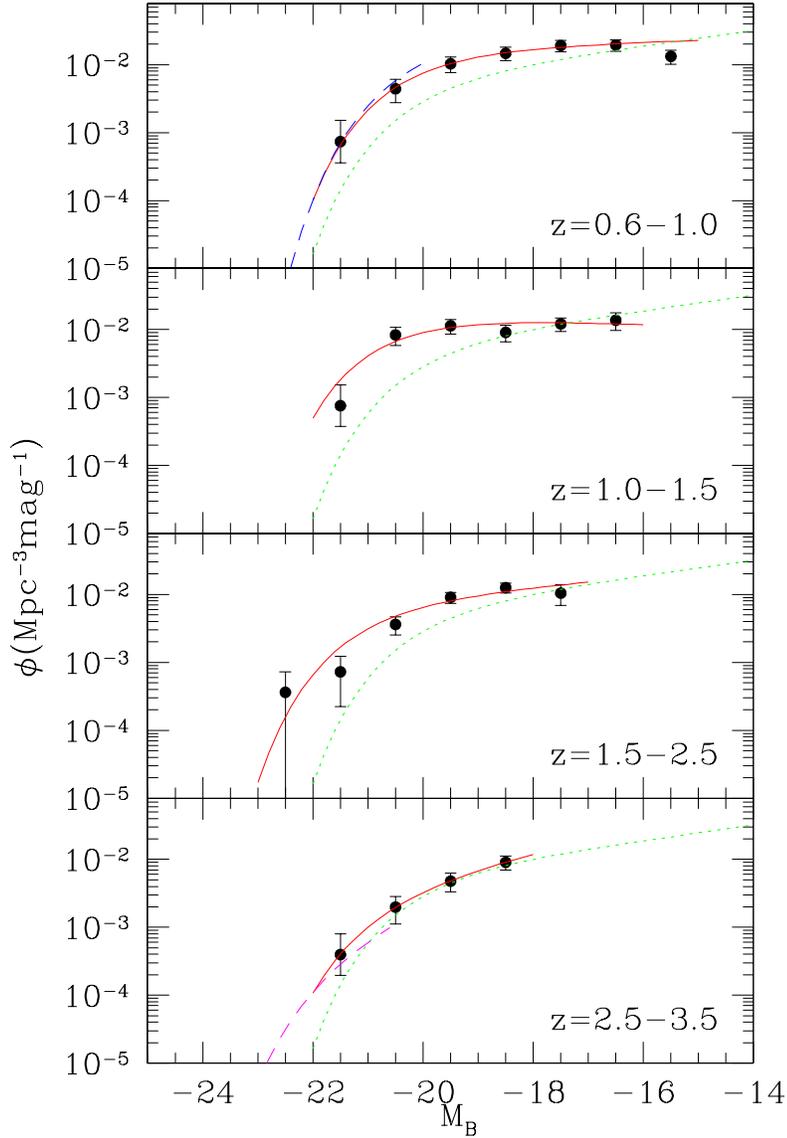}
\caption{Same as Figure~\ref{fig_lfk} for rest $B$-band LFs. The local $B$-band luminosity function denoted as dotted line is from \citet{bla01}, converted from their $g^*$ band to $B$ band. 
{\it Top}: $0.75<z<1.0$ rest $B$-band LF derived on CFRS \citep{lil95} is also superimposed as a dashed line.
{\it Bottom}: In the $2.5<z<3.5$ bin, a dashed line shows the rest $V$-band LF for Lyman-break galaxies at $\langle z\rangle =3.04$ by \citet{sha01}, converted to rest-$B$ band and $\Omega_m=1.0$, $h=0.75$ cosmology.
\label{fig_lfb}}
\end{figure}

The rest $B$-band LFs at $0.6<z \leq 3.5$ are shown in Figure~\ref{fig_lfb}, and the best-fit Schechter parameters are listed in Table~\ref{tbl_lf}.
Superposed dotted lines are the local $B$-band LFs from the SDSS \citep{bla01} converted from $g^*$-band.
In the $0.6<z \leq 1.0$ bin panel, we also superposed as a dashed line the $0.75<z<1.0$ rest $B$-band LF derived in the CFRS \citep{lil95} for the $\Omega_m=1$, $\Omega_\lambda=0$, and $h=0.75$ case.
Although CFRS samples only the bright end of the LF, the agreement between CFRS and our result for $0.6<z \leq 1.0$ is quite good, both show the brightening by $0.5$ mag from the local rest $B$ LF.
Note that CFRS is the $I$-band magnitude-limited sample and ours is $K^\prime$-band limited, though both of these bands trace above $4000$\AA~at $z<1.0$.

In the $2.5<z \leq 3.5$ panel of Figure~\ref{fig_lfb}, the dashed line represents the rest $V$-band LF of Lyman-break galaxies at $z\sim 3$ by \citet{sha01}.
We have converted their rest $V$-band magnitude to our rest $B$-band magnitude by using the rough estimate of average color of observed $(H-K)\sim0.40$, which approximately corresponds to the rest ($B-V$), at $z\sim 3$ derived on HDF-N galaxies by \citet{pap02}.
We have also reestimated the $M^*_V$ and $\phi^*$ values assuming a cosmology with $\Omega_m=1.0$ and $h=0.7$.
Though band conversion has some uncertainty ($\sim \pm 0.5$), our rest $B$ LF at $z\sim3$ is in good agreement with that of \citet{sha01} which is part of the rest UV spectroscopically selected sample of \citet{ste99}. 
The agreement of LFs at $z\sim3$ of our $K^\prime$-selected and their optically selected samples means that there are very few, if any, red candidates of old non-star-forming galaxies that would not be detected in optical bands but in near-infrared bands, which is also suggested by \citet{sha01} and \citet{pap02}.
On the contrary, we can conclude that our $K^\prime$-selected sample does not have a large deficit of young blue star-forming galaxies at least in brighter magnitudes.

In contrast to the rest $K^\prime$ LF, the rest $B$ LF shows brightning with epoch up to $z<2.5$.
The bright end of the LF appears to continue brightening by $0.4$mag from $0.6<z \leq 1.0$ to $1.0<z \leq 1.5$, and by a further $0.4$mag from $1.0<z \leq 1.5$ to $1.5<z \leq 2.5$, keeping the faint-end slope constant, though its brightening degree is within the $M^*$ error caused by photometric and redshift errors.
The bright-end amplitude of the rest $B$ LF at $1.5<z \leq 2.5$ clearly exceeds over the local $B$-band LF, while it declines at $2.5<z \leq 3.5$ almost to the local amplitude.

This brightening evolution in rest $B$ LF is also seen in \citet{saw97} for the HDF F814W$_{AB}<27$ sample.
They have found not only the brightening but also the steepening of the faint end slope with increasing redshift to $z=3$.
The difference in faint-end slope is presumably caused by the difference of the observed band used to construct the magnitude-limited sample, which will be discussed later.

In the rest $B$ band, we have found a strong luminosity evolution at $z<3$ which was not seen in the rest $K^\prime$ band.
In next subsection, we further investigate this trend in the bluer UV band, which directly relates to the star formation rate of a galaxy.

\subsection{Rest UV 2000\AA~Frame Luminosity Function}

We derive the rest UV LF at $2000$\AA~for which \citet{sul00} have derived the local LF for an ultraviolet selected sample.
We have evaluated photometric redshifts, as well as best-fit SEDs, for all the galaxies in our $K^\prime$-selected sample.
Therefore we can derive absolute magnitudes of any rest wavelength in the AB system, although the derived magnitudes would suffer directly the error of the photometric redshift evaluation.
Here we derive the absolute magnitude, $M_{2000}$ as follows,

\begin{equation}
M_{2000}=m_{AB}+2.5\mbox{log}(1+z)-5\mbox{log}(d_L(z)/10\mbox{pc}),
\end{equation}

where m$_{AB}$ is the observed redshifted magnitude at $2000(1+z)$\AA~and $d_L(z)$ is the luminosity distance.
It is to be noted that the derived rest UV luminosity is uncorrected for dust extinction, which is estimated to be in the range of $A_{1700}=0.0-4.0$mag at $z\sim3$ \citep{pap02}. 

\begin{figure}
\plotone{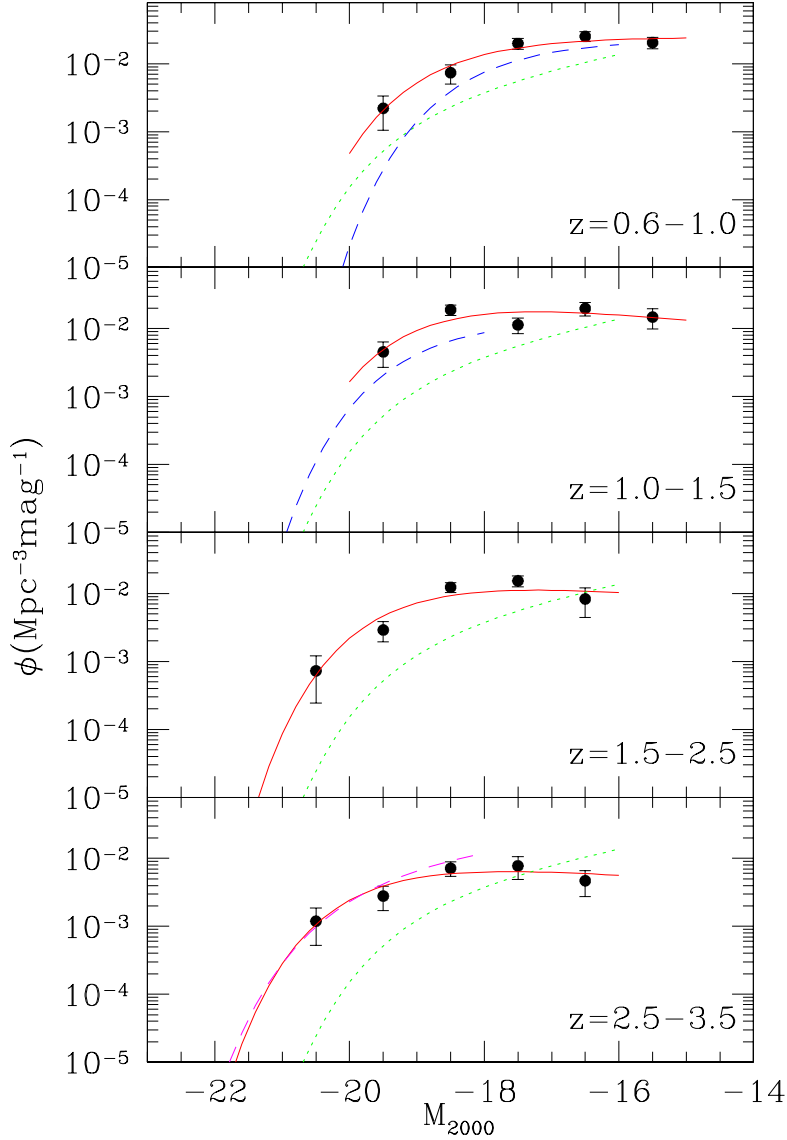}
\caption{Same as Figure~\ref{fig_lfk} for rest UV($2000$\AA~$AB$)-band LFs. 
The local $2000$\AA~luminosity function by \citet{sul00} for $h=0.75$ is denoted as dotted line converted from their magnitude system to AB system with a $2.29$ mag offset. 
In the bins for $0.6<z<1.0$ and $1.0<z<1.5$, dashed lines show the rest $2000$\AA~LF for $0.5<z<0.9$ and $1.0<z<1.5$ derived by \citet{cow99} for $\alpha=-1.0$ and $h=0.75$.
In the bin for $2.5<z<3.5$ panel, the dashed line shows the rest $1700$\AA~LF for Lyman-break galaxies at $\langle z\rangle =3.04$ by \citet{ste99}. 
\label{fig_lfu}}
\end{figure}

The rest $2000$\AA~LFs at $0.6<z \leq 3.5$ are represented in Figure~\ref{fig_lfu} and best-fit Schechter parameters are again listed in Table~\ref{tbl_lf}.
Superposed dotted lines are the local rest $2000$\AA~LF \citep{sul00} for $h=0.75$ converted to $AB$ system by a 2.29 mag offset \citep{cow99}.
As was seen in the rest $B$ band LF, the rest $2000$\AA~LFs also show a luminosity evolution that is stronger than in the rest $B$ LF and continues up to $2.5<z<3.5$ bin.
The $M_{2000}^*$ value is getting luminous from $z=0.6$ to $z=3.5$ over $1$ mag, which is more significant than $M_{2000}^*$ errors.

\citet{cow99} have derived the rest $2000$\AA~LF at $0.5<z<0.9$ and $1.0<z<1.5$.
We represent their LFs in the case of $\alpha=-1.0$ and $h=0.75$ as dashed line in corresponding redshift bins of Figure~\ref{fig_lfu}.
They mentioned that their value of ultraviolet luminosity density, which is lower than that of other studies, would partly be caused by field variation. 
Taking account of this normalization fluctuation, as well as their fairly large error of $\pm0.4$mag in determination of $M^*$, our results agree with their LFs.

For $2.5<z \leq 3.5$ in Figure~\ref{fig_lfu}, the dashed line represents the rest $1700$\AA~LF of Lyman-break galaxies at $\langle z\rangle=3.04$ by \citet{ste99}.
We ignore the rest-band difference between $2000$ and $1700$\AA~because the correction would be less than $\delta M\simeq 0.13$.
Again as seen in the rest $B$ LF, our estimate of the rest UV LF at $z\sim3$ is in good agreement with that of \citet{ste99}, who have estimated only the bright end ($M_{1700}<-20$).

As has been noticed in rest $B$ band LF, we can see less faint galaxies at higher redshift in this rest UV LF.
This is probably caused by the fact that our $K^\prime$-limited sample is insensitive to blue dwarf galaxies.
The deficit of faint-end galaxies in the $K$-selected sample is also seen in local universe.
The LFs of $K$-selected sample (e.g., \citealp{col01}; Table 3) have shallower faint-end slopes $\alpha \sim -1.0$ than those of optical-selected sample (e.g., \citealp{bla01}), $\alpha \sim -1.20$.


\section{Evolution of Rest UV-$K^\prime$ Color}

\begin{figure}
\plotone{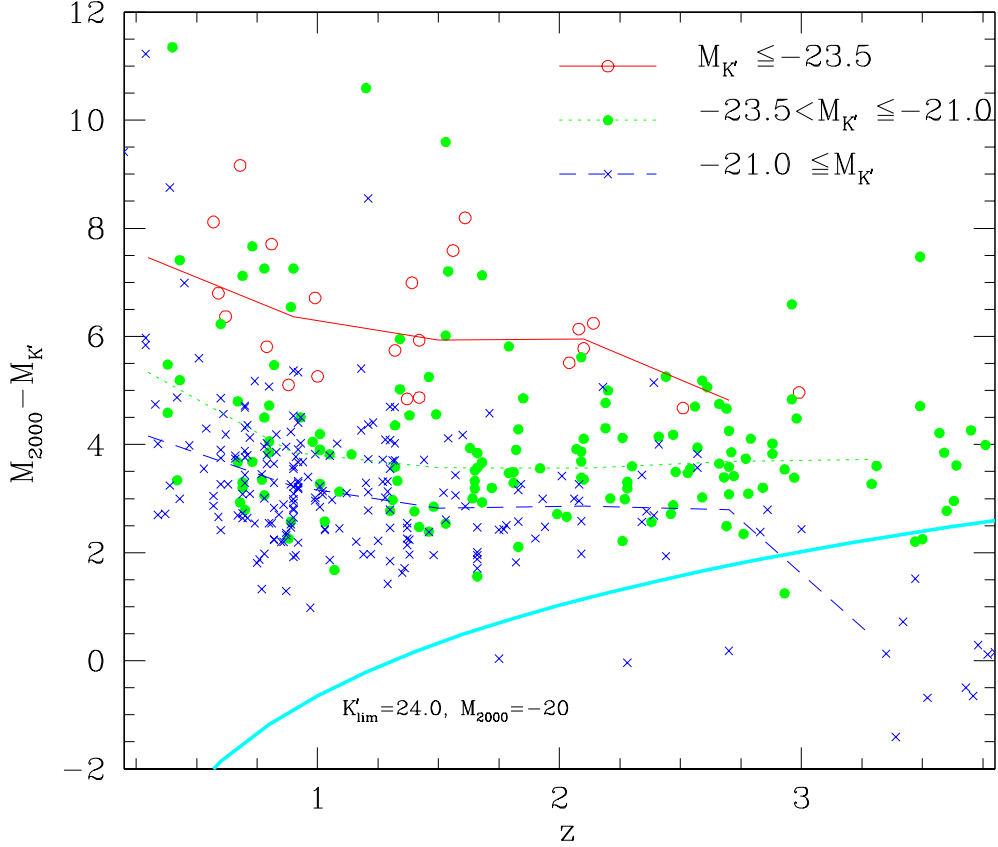}
\caption{Rest $M_{2000}-M_{K^\prime}$ color evolution with redshift.
Open circles, filled circles and crosses denote the subsamples with respect to the absolute $K^\prime$ magnitudes, $M_{K^\prime}\leq-23.5$, $-23.5<M_{K^\prime}\leq-21.0$ and $-21.0<M_{K^\prime}$, respectively.
The median $M_{2000}-M_{K^\prime}$ color for each redshift is traced by solid, dotted and dashed lines from bright subsample to faint subsample, respectively.
We show a thick solid line as a fiducial line denoting the lower color limit corresponding to our sample's magnitude limit of $K^\prime_{lim}=24.0$ galaxy for constant $M_{2000}=-20$ ignoring the small $k-$correction at $K^\prime$-band.
Galaxies brighter than $M_{2000}=-20$ would be detectable above this line.
If $M_{2000}=-19$ galaxy, you may shift the lower limit upper by $1$mag.
\label{fig_col}}
\end{figure}

In the previous section, concerned with rest-frame luminosity functions, we found the rest $K^\prime$-band LF has almost no evolution, while the $B$- and the UV-band LFs show drastic luminosity evolution; that is, galaxies get brighter in bluer bands with increasing redshift.
As a natural consequence, galaxies are inferred to be getting bluer in rest frame with redshift.
To see this trend more definitely, we show the rest $M_{2000}-M_{K^\prime}$ color evolution with redshift in Figure~\ref{fig_col}.
The rest-frame $2000$\AA~and $K^\prime$ luminosity of each galaxy selected in $K^\prime$ band can be calculated with the same manner described above.
Note that this color estimate is not based on the same aperture magnitudes in $2000$\AA~and $K^\prime$ bands.
We categorized the sample galaxies according to their absolute rest $K^\prime$-band magnitudes, $M_{K^\prime}$, which are fairly good indicators of the stellar mass.
We have mentioned in \S $4.1$ that $M_{K^\prime}$ values derived from the extrapolation of SED template might be a poor proxy of stellar mass at $z \sim3$, as suggested by \citet{sha01} and \citet{pap02}. 
Though predominant uncertainty caused by the second old population would increase the absolute stellar mass estimate systematically, the correlation between $M_{K^\prime}$ and stellar mass would be maintained if all the galaxies at $z\sim3$ are assumed to have similar mass ratios of the young to old stellar populations.
It was found that the derived stellar masses have a fairly tight correlation with extinction-uncorrected rest $V$ luminosity \citep{sha01}. 
\citet{pap02} have also found the correlation even with the rest UV luminosity with a scatter of $\sim0.3$dex, though \citet{sha01} have found only weak correlation.
However, we should note again that the extinction is still uncertain in the UV luminosity at $z \sim 3$.
In Figure~\ref{fig_col}, the most massive $M_{K^\prime}\leq-23.5$ sample is represented by open circles.
The $-23.5<M_{K^\prime}\leq-21.0$ and $-21.0<M_{K^\prime}$ samples are represented by filled circles and crosses, respectively.
The median $M_{2000}-M_{K^\prime}$ color for each redshift is traced by solid, dotted and dashed lines from the most massive sample to the least massive one, respectively.

Two significant trends can be seen in this figure.
First, all three subsamples have a mild bluing trend in $M_{2000}-M_{K^\prime}$ with increasing redshift as expected from our results on rest LFs.
For example, $M_{K^\prime}\leq-23.5$ sample has median color $M_{2000}-M_{K^\prime}=7.46$ around $z=0.3$, while at $z=2.7$ there are no such red galaxies and median color becomes as blue as $4.73$.
The bluing trend we found has already been indicated by several studies (e.g., \citealp{cow96,pap02}) and is as qualitatively expected from a passive evolution model.
\citet{pap02} suggests that the UV-optical rest-frame SEDs of the Lyman break galaxies at $2.0 \lesssim z \lesssim 3.5$ are much bluer than those of present-day spiral and elliptical galaxies.
The other two subsamples, $-23.5<M_{K^\prime}\leq-21.0$ and $-21.0<M_{K^\prime}$ also show this bluing trend, at least at $z<1.5$, and almost no color changes have been found beyond $z=1.5$.
Our sample is so small that we cannot discriminate the difference of rapidity of this bluing trend among subsamples.

It should be noted that an observational limit might affect this diagram.
It is difficult to show explicitly the observational limit of $K^\prime \leq 24.0$ in this rest-frame plane.
For simplicity, we show in Figure~\ref{fig_col} a fiducial line (heavy solid line) denoting the bluer color limit of a galaxy having constant luminosity of $M_{2000}=-20$, which can be observed in the $K^\prime$-band image ($K^\prime \leq 24.0$), ignoring the small $k-$correction (e.g., \citealp{man01}) at the $K^\prime$ band.
Galaxies brighter than $M_{2000}=-20$ would be detectable above this line.
For a galaxy with $M_{2000}=-19$, you may shift the lower limit upward by $1$mag.
At higher redshift, it is found that galaxies that are bluer and fainter in $2000$\AA~are not likely to stay in this $K^\prime$-limited sample.
Our subsample of $-23.5<M_K \leq -21.0$ and $-21.0<M_K$ would have been affected by this observational bias which may cause a lack of blue galaxies at high redshift if such blue population really exists.
The blueing trend of these subsample seems to stop at $z=1.5$ in Figure~\ref{fig_col}, though it may be possibly attributed to this bias.

On the other hand, for $M_K \leq -23.5$ subsample, which seems to be less affected by this bias, all the galaxies become redder with time in $z<1$, while there are galaxies around $z \simeq 1$ as blue as those at $z \simeq 3$.
It seems that the scatter of the color becomes larger in the nearby universe.
The same trend can be seen in the rest-frame $U-B$ color of the brightest ($M_B<-20.3$) galaxy sample in HDF by \citet{dic00}.
These galaxies with similar stellar masses keep their blue color throughout the passive evolution during a longer timescale than the typical timescale of an intense starburst event.
This implies that a merging process (e.g., \citealp{lac93}) and subsequent star formation have been taking place continuously during $1 \lesssim z \lesssim 3$ epoch.
These blue galaxies of $M_K \leq -23.5$ at $z=1$ seem to be assembled from lower mass galaxies via. merging and are ongoing star formation.

Second, it is found that more massive galaxies tend to be redder in rest $M_{2000}-M_{K^\prime}$ color at every epoch.
The median color of massive galaxies is always redder than that of less massive galaxies.
The trend that red galaxies are more luminous than blue ones has already been established in the local universe for example by \citet{bla01}, who found that on average high surface brightness, red, and highly concentrated galaxies are more luminous than low surface brightness, blue, and less concentrated galaxies.
If rest $M_{2000}-M_{K^\prime}$ color indicates the age of a galaxy, the massive galaxies tend to have older, which can be easily inferred from the context of downsizing hypothesis, in which massive (spheroidal) galaxies formed at high z, followed at later epochs by the formation of less massive galaxies.



\section{Contribution to The Global Star Formation Rate Density}

In the last few years, there have been great advances in understanding the evolution of the global star formation rate density (SFRD).
At low redshift between $z=0$ and $z=1.5$, the SFRD is found to be steadily increasing with redshift.
At high redshift, some results (e.g., \citealp{mad98}) meet the SFRD peak around $z\sim1.5$, while others (e.g., \citealp{ste99}) maintain it relatively constant.
The estimate of SFRD based on the ultraviolet luminosity of galaxies is strongly affected not only by dust obscuration, which many investigators took into account but also by the faint end of the LF.
At last in this study, we derive the SFRD based on our evaluated rest UV LF.

The SFRD can be estimated by integrating over the best-fit Schechter function to infinity for the rest UV LF in each redshift bin.
Equation (4) below was used to convert luminosity at $2000$\AA, $L_{2000}$ to an SFR based on the calibration given by \citet{mad98} for a Salpeter IMF,

\begin{equation}
SFR(M_\odot \mbox{yr} ^{-1}) = 1.26\times10^{-28}L_{2000}(\mbox{ergs}~ \mbox{s}^{-1} \mbox{Hz}^{-1}).
\end{equation}

\begin{figure}
\plotone{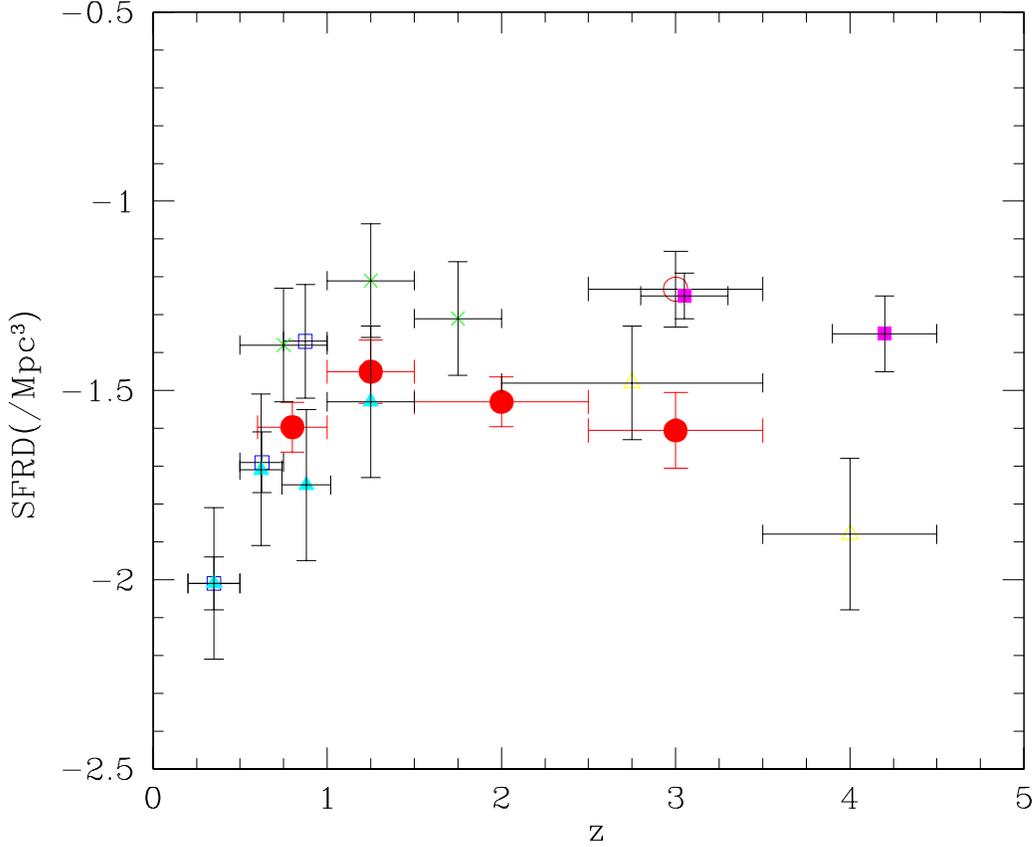}
\caption{Star formation rate density as a function of redshift (Madau plot) derived from the integration of UV-LF. 
The estimates of this work are shown in filled circles. 
The different data are taken from previous estimates of Lilly et al. (1996; {\it open squares}), Connoly et al. (1997; {\it crosses}), Cowie et al. (1999; {\it filled triangles}), Madau et al. (1998; {\it open triangles}), and Steidel et al. (1999; {\it filled squares}).
The dust extinction is uncorrected for in all the estimates.
If we change our rest UV LF slope from $\alpha=-0.83$ to $\alpha=-1.6$ as \citet{ste99} have adopted at $z=3$, the estimate at $z=3$ shifts up to the open circle from the filled circle.
\label{fig_sfr}}
\end{figure}

The result is represented in Figure~\ref{fig_sfr}, in which several previous estimates are also plotted.
All the previous results except \citet{ste99} were derived by integrating the Schechter function to infinity.
For consistency, we increased the \citet{ste99} points by an additional factor $1.7$ as they have noted in the case of integration to infinity.
We checked our data point by recalculating the SFRD from other rest UV wavelengths such as $1500$ and $2800$\AA~as was adopted in previous studies and found that our SFRD estimate is stable within a fluctuation of $0.2$.
The dust extinction is uncorrected for in all the estimates.

Our results are denoted as solid circles and are broadly consistent with others at $z<1.5$, though they show slight deficiencies at $z>1.5$, where there are yet no established observations.
This has probably arisen because our estimated rest UV LF of the $K^\prime$-selected sample has a shallower faint-end slope than that of \citet{ste99}.
For an example, if we change the LF slope at $2.5<z \leq 3.5$ from $\alpha=-0.83$ to $\alpha=-1.6$ as \citet{ste99} have adopted at $z=3$, our estimate at $z=3$ in Figure~\ref{fig_sfr} shifts up to the open circle, which is almost consistent with \citet{ste99}.
Otherwise, if the slope of the rest UV LF at $z=3$ is really as steep as $\alpha=-1.6$, our $K^\prime<24.0$ limited sample contributes only $42\%$ of the total UV luminosity density, that is, of the total star formation rate density at $z=3$.
The remaining $58\%$ should be charged to faint galaxies with $M_{2000}>-19$ for which our sample seems to have a deficiency of galaxies compared with the LF derived by \citet{ste99}.
These missed faint galaxies correspond to very blue galaxies, which are probably affected by the observational limit in $K^\prime$ band as has been shown in \S 5.
However, we should note here that as they have mentioned the $\alpha$ derived by \citet{ste99} has a large uncertainty because it is determined only at the bright end ($M_{1700}<-20$) of the LF.
\citet{pol01} have estimated the photometric redshifts for $I$-selected galaxies and obtained a shallower faint end slope $\alpha=-1.37\pm0.20$ for the rest $1700$\AA~LF of $2.5<z \leq 3.5$ galaxies down to $M_{1700}=-19$.
The existence of a substantial population of such faint ($M_{1700}>-20$) and blue Lyman break galaxies at $z>3$ is still questionable.



\section{Summary}

We have constructed a very deep $K^\prime$-selected multicolor $BVRIz^\prime JK^\prime$ sample of field galaxies on the Subaru Deep Field.
Based on this multicolor data, a photometric redshift for each sample galaxy was estimated.
Our principal conclusions are the following:

\begin{enumerate}

\item The overall N(z) distribution for the $K^\prime \leq 21$ sample is consistent with previous observations, and we for the first time derived the N(z) down to $K^\prime=24$.
Taking account of dust extinction and selection effects of the sample, the observed N(z) distribution at the most secure magnitude bin, $22<K^\prime \leq 23$, is well described with the PLE model, while a recent version of the hierarchical galaxy formation model shows an apparent deficiency of galaxies, especially at $z>2$.
The PLE model with realitic dust evaluation can account for the abundance of the high z ($z>2$) galaxies.
In the faintest bin of N(z), $23<K^\prime \leq 24$, both of the model predictions are found to be inconsistent with the observations.
The discrepancy in the faintest bin is presumably caused by the photometric errors that give an incorrect solution of photometric redshifts. 

\item The rest $K^\prime$, $B$, and UV($2000$\AA) band luminosity functions and their evolution have been investigated.
The rest $K^\prime$-band luminosity function shows almost no evolution up to $z=3$. 
The rest $B$ luminosity function shows mild brightning with epoch up to $z=2.5$, though its brightening degree is within the $M^*$ error caused by photometric and redshift errors.
The rest $2000$\AA~LFs also show a luminosity evolution that is more significant than in the rest $B$ LF and continues up to $z=3.5$.
The $M_{2000}^*$ value is getting luminous from $z=0.6$ to $z=3.5$ over $1$ mag, which is more significant than $M_{2000}^*$ errors.

\item The evolutionary trend of rest luminosity functions appears to correspond to the bluing trend of the rest $UV-K^\prime$ color with increasing redshift.
We also found that more massive galaxies are redder at every epoch.

\item The findings about the evolution of rest luminosity functions are qualitatively in favor of the PLE model. 
The star formation activity in a galaxy changes the rest blue luminosity, while the amount of old stellar population does not change.
The resulting rest $K^\prime$-band luminosity function showing almost no evolution up to $z=2.5$ can also be consistent with the hierarchical galaxy formation model if $M/L_K$ decreases with redshift; that is, the number density of $K^\prime$-band luminosity-selected galaxies does not significantly change with redshift, while the number density of stellar mass-selected galaxies decreases with look-back time. 


\item We have evaluated the contribution to the cosmic star formation rate density of our $K^\prime$-selected galaxies.
The star formation rate density derived from integration of our rest UV luminosity function is about $42\%$ of that derived from Lyman break galaxies at $z=3$.
The result suggests that a large fraction of global star formation rate density at $z>2$ could represent the contribution of faint blue galaxy population, which has not yet explicitly been identified in a high z universe.

\end{enumerate}

Though the present results rely heavily on an availability of photometric redshifts, we are confident that we have successfully derived a coherent picture of evolution of $K^\prime$, $B$, and UV rest-frame luminosity functions over the range $0.6<z \leq 3.5$ to the faint end below the limit of the today's spectroscopic studies.
Our result raises the question of how such massive galaxies have formed at high redshift with efficient star formation, which is apparently consistent with the PLE picture in the $K$ band, within the framework of the structure formation in the CDM universe.
Of course, deeper NIR imaging surveys in future will allow us to derive more secure results for N(z) at the faintest magnitude and on the faint end of the rest LFs, in which we have obtained robust results in this study.
Deep observation at longer wavelength with $SIRTF$ and $ASTRO$-$F$ will be also required to constraint the SEDs of galaxies at $z>3$.
The technique that we have adopted in this study will be applied to future wide-field $K$-band surveys to determine the rest luminosity at any wavelength at any redshift down to the faintest magnitude.
On the other hand, ongoing large multiobject spectroscopic surveys are of course extremely important to reveal the galaxy evolution more reliably.

\acknowledgments

We deeply appreciate the devoted support of the Subaru Telescope staff for this project.
We thank the referee for helpful comments that improved the manuscript.
The research is supported the Ministry of Education, Science and Culture through Grant-in-Aid 11691140.

\clearpage


\clearpage 






\clearpage

\begin{deluxetable}{crrr}
\tablecaption{Summary of Optical Imaging Data for SDF. \label{tbl-1}}
\tablewidth{0pt}
\tablehead{
\colhead{band} & \colhead{exp.time}  & \colhead{seeing size} & \colhead{$m_{lim}$\tablenotemark{a}} \\
 & (sec) & (arcsec) & (mag) 
}
\startdata
$B$ & $9000$  & $0.6$ & $28.08$ \\
$V$ & $3600$  & $0.8$ & $26.28$ \\
$R$ & $14400$ & $0.7$ & $27.88$ \\
$I$ & $3600$  & $0.7$ & $25.93$ \\
$z^\prime$    & $3600$ & $0.8$ & $24.27$ \\
\enddata

\tablenotetext{a}{The $5\sigma$ limiting magnitudes in the Vega system within $2^{\prime \prime}$ aperture}

\end{deluxetable}

\begin{deluxetable}{lcccc}
\tablecaption{Chance Probability and $\chi^2$ (in parentheses) Values Obtained from Kolmogorov-Smirnov Test for N(z) Distribution. \label{tbl-ks}}
\tablewidth{0pt}
\tablehead{
\colhead{models} & \colhead{$19<K^\prime \leq 21$}  & \colhead{$21<K^\prime \leq 22$} & \colhead{$22<K^\prime \leq 23$} & \colhead{$23<K^\prime \leq 24$}
}
\startdata
PLE      & 1.97\%(8.90) & 1.03\%(10.2)   & 18.2\%(4.67)    & 0.232\%(13.3) \\
hierarchical & 9.32\%(5.91) & 76.4\%(1.72)   & 6.46e-4\%(24.7) & 4.49e-13\%(66.2)\\
\enddata

\end{deluxetable}


\begin{deluxetable}{lllll}
\tablecaption{The Best-Fit Shechter Parameters of Rest-Frame Luminosity Functions in Each Redshift Bin Derived on SDF \label{tbl_lf}}
\tablewidth{0pt}
\tablehead{
\colhead{Rest-frame} & \colhead{$z$ range}  & \colhead{$\alpha$} & \colhead{$M^*$} & \colhead{$\phi^*$} 
}
\startdata
$K^\prime$ 	& $0.6-1.0$ & $-1.35^{+0.05}_{+0.02}$ & $-25.03^{+0.12}_{-0.27}$ & $3.87^{+2.12}_{-0.41}\times10^{-3}$ \\
		& $1.0-1.5$ & $-1.35^{-0.01}_{-0.10}$ & $-25.02^{+0.49}_{-0.05}$ & $3.87^{+0.94}_{-2.18}\times10^{-3}$ \\
		& $1.5-2.5$ & $-1.37^{+0.09}_{+0.00}$ & $-24.95^{+0.22}_{+0.01}$ & $3.13^{+2.09}_{+0.09}\times10^{-3}$ \\
		& $2.5-3.5$ & $-1.70^{+0.17}_{-0.06}$ & $-23.95^{+0.11}_{-0.35}$ & $2.57^{+1.88}_{-1.75}\times10^{-3}$ \\
\tableline
$B$ 		& $0.6-1.0$ & $-1.07^{+0.01}_{-0.07}$ & $-20.26^{-0.26}_{-0.63}$ & $1.76^{-0.14}_{-0.67}\times10^{-2}$ \\
		& $1.0-1.5$ & $-0.92^{+0.13}_{-0.10}$ & $-20.61^{+0.19}_{+0.31}$ & $1.80^{+0.30}_{-0.70}\times10^{-2}$ \\
		& $1.5-2.5$ & $-1.19^{+0.25}_{+0.01}$ & $-21.10^{+0.37}_{+0.01}$ & $8.31^{+7.82}_{+1.76}\times10^{-3}$ \\
		& $2.5-3.5$ & $-1.49^{+0.45}_{-0.01}$ & $-20.82^{+0.44}_{-0.30}$ & $3.86^{+6.84}_{-3.49}\times10^{-3}$ \\
\tableline
$U$($2000$\AA) 	& $0.6-1.0$ & $-0.99^{+0.34}_{-0.11}$ & $-18.49^{+0.31}_{-0.11}$ & $2.79^{+0.80}_{-0.08}\times10^{-2}$ \\
		& $1.0-1.5$ & $-0.75^{+0.10}_{-0.04}$ & $-18.71^{+0.07}_{-0.06}$ & $3.49^{-0.08}_{-0.76}\times10^{-2}$ \\
		& $1.5-2.5$ & $-0.83^{+0.21}_{-0.04}$ & $-19.12^{+0.18}_{-0.05}$ & $1.95^{+0.88}_{+0.36}\times10^{-2}$ \\
		& $2.5-3.5$ & $-0.83^{+0.26}_{-0.13}$ & $-19.54^{+0.23}_{-0.12}$ & $1.11^{+0.67}_{-0.04}\times10^{-2}$ \\
\enddata

\end{deluxetable}

\clearpage


\begin{thebibliography}{}
\bibitem[Arimoto \& Yoshii (1987)]{ari87} Arimoto,N., \& Yoshii,Y. 1987, \aap, 173, 23
\bibitem[Arimoto, Yoshii \& Takahara (1992)]{ari92} Arimoto,N., Yoshii,Y., \& Takahara,F. 1992, \aap, 253, 21
\bibitem[Barger et al.(1999)]{bar99} Barger,A.J., Cowie,L.L., Trentham,N., Fulton,E., Hu,E.M., Songaila, A., \& Hall,D. 1999, \aj, 117, 102
\bibitem[Bertin \& Arnouts(1996)]{ber96} Bertin,E., \& Arnouts,S. 1996, \aaps, 117, 393
\bibitem[Bolzonella, Miralles, \& Pello(2000)]{bol00} Bolzonella,M.J., Miralles,M., \& Pello,R. 2000, \aap, 363, 47
\bibitem[Blanton et al.(2001)]{bla01} Blanton,M.R., et al. 2001, \aj, 121, 2358
\bibitem[Bruzal \& Charlot(1993)]{bc93} Bruzal,G., \& Charlot,S. 1993, \apj, 405, 538
\bibitem[Calzetti etal.(2000)]{cal00} Calzetti, D., Armus, L., Bohlin, R.C., Kinney, A.L., Koornneef, J., \& Storchi-Bergmann, T., 2000, \apj, 533, 682
\bibitem[Cimatti et al.(2002)]{cim02} Cimatti, A. et al. 2002, \aap, 391, L1
\bibitem[Cole et al.(2001)]{col01} Cole,S., et al. 2001, \mnras, 326, 255
\bibitem[Connolly et al.(1997)]{con97} Connolly,A.J., Szalay,A.S., Dickinson,M.E., SubbaRao,M.U., \& Brunner, R.J. 1997, \apj, 486, L11
\bibitem[Cowie, Songaila, \& Barger(1999)]{cow99} Cowie,L.L., Songaila,A, \& Barger,A.J. 1999, \aj, 118, 603
\bibitem[Cowie et al.(1996)]{cow96}  Cowie,L.L., Songaila,A, Hu,E.M., \& Cohen,J.G. 1996, \aj, 112, 839
\bibitem[Dickinson(2000)]{dic00} Dickinson,M. 2000 in {\it Building Galaxies: From the Primordial Universe to the Present}, Proceedings of the XIXth Moriond Astrophysics Meeting, eds. F. Hammer, T.X. Thuan, V. Cayatte, B. Guiderdoni, \& J. Tranh Than Van, (Paris: Ed. Frontieres), p. 257
\bibitem[Drory et al.(2001)]{dro01} Drory,N., et al., 2001, \apj, 562, L111
\bibitem[Ellis et al.(1996)]{eli96} Ellis,R.S., Colless,M., Broadhurst,T., Heyl,J., \& Glazebrook,K. 1996, \mnras, 280, 235
\bibitem[Felten(1977)]{fel77} Felten,J.E. 1977, \aj, 82,861
\bibitem[Folkes et al.(1999)]{fol99} Folkes,S., et al. 1999, \mnras, 308, 459
\bibitem[Fontana et al.(2000)]{fon00} Fontana,A., D'Odorico,S., Poli,F., Giallongo,E., Arnouts,S., Cristiani,S., Moorwood,A., \& Saracco,P. 2000, \aj, 120, 220
\bibitem[Fontana et al.(1999)]{fon99} Fontana,A., Menci,N., D'Odorico,S., Giallongo,E., Poli,F., Cristiani,S., Moorwood,A., \& Saracco,P. 1999, \mnras, 310, L27
\bibitem[Furusawa et al.(2003)]{fur03} Furusawa,H., et al. 2003 (Paper IV) in preparation.
\bibitem[Kashikawa et al.(2002)]{kas02} Kashikawa,N., et al.  2002, PASJ, in press
\bibitem[Kauffmann \& Charlot(1998)]{kc98} Kauffmann,G., \& Charlot, S. 1998, \mnras, 297, L23 (KC98)
\bibitem[Lacey \& Cole(1993)]{lac93} Lacey,C., \& Cole,S. 1993, \mnras, 1993, 262, 627
\bibitem[Lowenthal et al.(1997)]{low97} Lowenthal,J.D., et al. 1997, \apj, 481, 673
\bibitem[Lilly et al.(1996)]{lil96} Lilly,S.J., Le F\'evre,O., Hammer,F., \& Crampton,D. 1996, \apj, 460, L1
\bibitem[Lilly et al.(1995)]{lil95} Lilly,S.J., Tresse,L., Hammer,F., Crampton,D., \& Le F\'evre,O. 1995, \apj, 455, 108
\bibitem[Loveday(2000)]{lov00} Loveday,J. 2000, \mnras, 312, 517
\bibitem[Maihara et al.(2001)]{mai01} Maihara,T., et al.  2001, PASJ, 53, 25 (Paper I)
\bibitem[Madau, Pozzetti, \& Dickinson(1998)]{mad98} Madau,P., Pozzetti,L., \& Dickinson,M. 1998, \apj, 498, 106
\bibitem[Madau (1995)]{mad95} Madau,P., 1995, \apj,  441, 18
\bibitem[Mannucci et al.(2001)]{man01} Mannucci,F., Basile,F., Poggianti,B.M., Cimatti,A., Daddi,E., Pozzetti,L., \& Vanzi,L. 2001, \mnras, 326, 745
\bibitem[Miyazaki et al.(1998)]{miy98} Miyazaki,S., Sekiguchi,M., Imi,K., Okada,N., Nakata,F., \& Komiyama,Y. 1998, SPIE, 3355, 363
\bibitem[Nagashima et al.(2003)]{nag03} Nagashima,M., Yoshii,Y., Totani,T., \& Gouda,N. 2003, \apj, in press.
\bibitem[Nagashima et al.(2001)]{nag01} Nagashima,M., Totani,T., Gouda,N., \& Yoshii,Y. 2001, \apj, 557, 505
\bibitem[Ouchi et al.(2003)]{ouc03} Ouchi, M.~et al. 2003, \apj, in press (Paper II)
\bibitem[Papovich, Dickinson, \& Ferguson(2002)]{pap02} Papovich,C., Dickinson,M., \& Ferguson,H.C. 2002, \apj, 559, 620
\bibitem[Poli et al.(2001)]{pol01} Poli,F., Menci,N., Giallongo,E., Fontana,A., Cristiani,S., \& D'Odorico,S. 2001, \apj, 551, L45
\bibitem[Rudnick et al.(2001)]{rud01} Rudnick et al. 2001, \aj, 122, 2205
\bibitem[Saracco et al.(2001)]{sar01} Saracco,P., et al. 2001, \aap, 375, 1
\bibitem[Sawicki, Lin \& Yee(1997)]{saw97} Sawicki,M.J., Lin,H., \& Yee,K.C. 1997, \aj, 113, 1
\bibitem[Schlegel, Flinkbeiner \& Davis(1998)]{sch98} Schlegel,D.J., Finkbeiner,D.P., \& Davis,M. 1998, \apj, 500, 525
\bibitem[Shapley et al.(2001)]{sha01} Shapley,A.E., Steidel,C.C., Adelberger,K.L., Dickinson,M., Giavalisco,M., \& Pettini,M. 2001, \apj, 562, 95
\bibitem[Steidel et al.(1999)]{ste99} Steidel,C.C., Adelberger,K.L., Giavalisco,M., Dickinson,M., \& Pettini,M. 1999, \apj, 519, 1
\bibitem[Steidel et al.(1996)]{ste96} Steidel,C.C., Giavalisco,M., Dickinson,M., \& Adelberger,K.L. 1996, \aj, 112, 352
\bibitem[Sullivan et al.(2000)]{sul00} Sullivan,M., Treyer,M.A., Ellis,R.S., Bridges,T.J., Milliard,B., \& Donas,J. 2000, \mnras, 312, 442
\bibitem[Totani et al.(2001a)]{tot01a} Totani,T., Yoshii,Y., Iwamuro,F., Maihara,T., \& Motohara,K. 2001a, \apj, 550, L137
\bibitem[Totani et al.(2001b)]{tot01b} Totani,T., Yoshii,Y., Iwamuro,F., Maihara,T., \& Motohara,K. 2001b, \apj, 559, 592 
\bibitem[Totani \& Yoshii(2000)]{ty00} Totani, T. \& Yoshii, Y. 2000, \apj, 540, 81
\bibitem[Yoshii(1993)]{yos93} Yoshii, Y. 1993, \apj, 403, 552
\end{thebibliography}
\end{document}